\documentclass[amssymb,prb,twocolumn,floats,amsmath,superscriptaddress]{revtex4}

\usepackage{graphicx}

\begin{document}


\title{High-reflectivity broadband distributed Bragg reflector lattice matched to ZnTe}

%
\author{W.~Pacuski}
\email{Wojciech.Pacuski@fuw.edu.pl}
\affiliation{Institute of Solid State Physics, University of Bremen, P.O. Box 330-440, D-28334 Bremen, Germany}
\affiliation{Institute of Experimental Physics, University of Warsaw,
Ho\.za 69, PL-00-681 Warszawa, Poland}

\author{C. Kruse}
\affiliation{Institute of Solid State Physics, University of Bremen, P.O. Box 330-440, D-28334 Bremen, Germany}

\author{S. Figge}
\affiliation{Institute of Solid State Physics, University of Bremen, P.O. Box 330-440, D-28334 Bremen, Germany}

\author{D. Hommel}
\affiliation{Institute of Solid State Physics, University of Bremen, P.O. Box 330-440, D-28334 Bremen, Germany}


\begin{abstract}
We report on the realization of a high quality distributed Bragg reflector with both high and low refractive index layers lattice matched to ZnTe. Our structure is grown by molecular beam epitaxy and is based on binary compounds only. The high refractive index layer is made of ZnTe, while the low index material is made of a short period triple superlattice containing MgSe, MgTe, and ZnTe. The high refractive index step of $\Delta n=0.5$ in the structure results in a broad stopband and the reflectivity coefficient exceeding 99\% for only 15 Bragg pairs.

\end{abstract}

\keywords{distributed Bragg reflector, II-VI semiconductors, zinc compounds,  magnesium compounds, tellurides,  molecular beam epitaxy, semiconductor growth, X-ray diffraction,  refractive index, reflectivity spectrum}

\pacs{42.70.Qs, 42.79.Fm, 78.20.Ci, 78.66.Hf, 81.15.Hi}


\maketitle


ZnTe-based heterostructures such as CdTe/ZnTe quantum dots (QDs) \cite{karczewski99,tinjod03} and quantum wells \cite{Ueta02,Ochiai02,Nomura04,Nomura06} exhibit emission in the green to orange spectral range, where devices based on III-V compounds have a low quantum efficiency. This spectral range is of particular interest for data communication using plastic optical fibres  since  they exhibit a minimum of the light absorption in that region.\cite{Daum02} Highly efficient light sources like monolithic vertical-cavity surface-emitting lasers (VCSELs) and resonant-cavity light emitting diodes radiating in this range can be realized based on the CdTe/ZnTe system if high reflectivity distributed Bragg reflectors (DBRs) lattice matched to a ZnTe substrate are available. Furthermore, by incorporation of CdTe QDs containing single Mn ions into a high-quality microcavity, applications concerning spintronics and quantum information science might be possible.\cite{Gall09,Goryca09}


A DBR is based on alternating layers of high and low refractive index. Basically a DBR can be made of any transparent materials but it should be made of semiconductors if is combined with semiconductor heterostructures in order to create high quality optoelectronic devices. Therefore, the materials used within the DBR should have the same lattice parameter in order to avoid defects like misfit dislocations. The requirements of the lattice matching and of a high refractive index step make the design and fabrication of those DBRs quite challenging. A number of II-VI ternary and quaternary compounds have been examined\cite{Peiris01} in order to find good materials for DBRs working in the visible spectral range. Creation of lattice matched DBRs based on II-VI compounds has been reported so far for three substrates: Cd$_{0.88}$Zn$_{0.12}$Te,\cite{Dang98} GaAs,\cite{Kruse04,Sebald09} and InP.\cite{Guo00,Tamargo01} Here we report the realization of a high-reflectivity DBR lattice matched to ZnTe.


The standard approach for the growth of DBR structures includes ternary alloys with composition adjusted to keep the same lattice parameter for the layers with high and low refractive index. For example Le~Si~Dang et al.\cite{Dang98} developed a II-VI DBR on Cd$_{0.88}$Zn$_{0.12}$Te substrate, with Cd$_{0.40}$Mg$_{0.60}$Te as the low refractive index layer and Cd$_{0.75}$Mn$_{0.25}$Te as the high refractive index layer. All three materials have different energy gaps and refractive indices but the same lattice parameter, therefore relaxation is not observed even after growth of tens of DBR pairs. Such an approach is quite difficult to apply in the case of a DBR lattice matched to ZnTe.

\begin{figure}[bt]
\includegraphics[width=1\linewidth,clip]{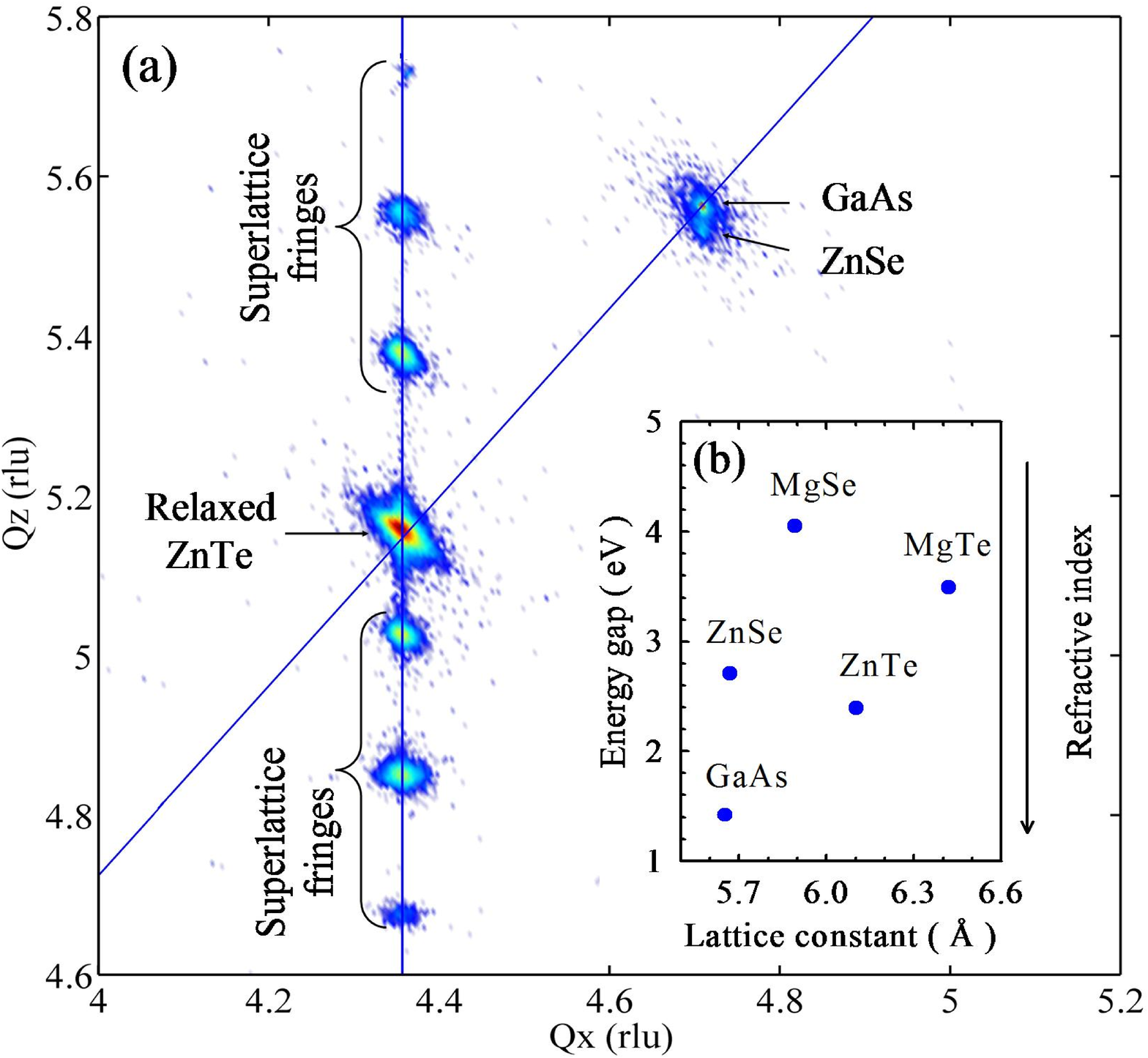}
\centering \caption[]{(color online). (a) Reciprocal space mapping close to (335) X-ray reflex. Qy is in layer plane, Qz is parallel the growth axis, both are in reciprocal lattice unit (rlu) with a dimension of (1/{\AA}). The line representing relaxed materials crosses the spots of ZnTe and GaAs. The vertical line connecting the superlattices fringes crosses the spot of ZnTe. This shows that the superlattice is pseudomorphic to the relaxed ZnTe buffer. (b) Energy gap versus lattice parameter for semiconductors used in this work. Note that refractive index increases, when energy gap of semiconductor decreases.}
\label{fig:XRD}
\end{figure}

The choice of ZnTe with a refractive index above $n=3$ in the visible spectral range as the high refractive index layer is straightforward. Unfortunately, there is no binary or ternary II-VI compound which could act as low refractive index layer with zinc blende structure and lattice parameter of ZnTe [see Fig.~\ref{fig:XRD}(b)]. One possibility is the use of a quaternary compound (Zn,Mg)(Te,Se)\cite{Ueta02,Nomura04} with well chosen composition. However, growth of thick structures with such a complex material while keeping the exact composition is comparatively difficult. If the Mg content is exceeding a certain limit, the zinc blende structure becomes unstable because MgTe and MgSe naturally crystallize in rock salt structure. On the other hand, for a low Mg content the refractive index step between (Zn,Mg)(Te,Se) and ZnTe will be low resulting in a DBR with a small stopband width. It has been shown that a short period double superlattice containing ternary compounds instead of quaternary compound can be grown in good quality.\cite{Ochiai02} Our approach is the use of a short period triple superlattice (SL) which allows us to avoid not only quaternary, but even ternary compounds. Moreover, the advantage of using superlattices for growth of II-VI DBRs is that one can use a relatively high Mg content in order to achieve a high refractive index step, and consequently, a broad DBR stopband width.\cite{Tawara00,Kruse04,Sebald09}

\begin{figure}[bt]
\includegraphics[width=1\linewidth,clip]{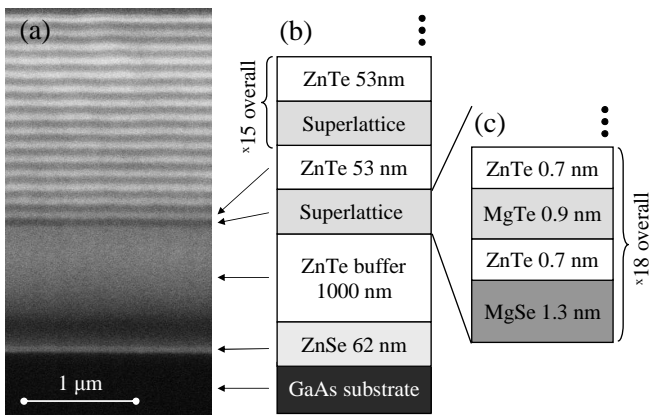}
\centering \caption[]{Structure of distributed Bragg reflector lattice matched to ZnTe buffer. (a) Cross-section image obtained with scanning electron microscope. (b) Scheme of layers. Material with high refractive index is ZnTe. Material with low refractive index is a short period triple superlattice MgSe/ZnTe/MgTe/ZnTe shown in (c). Layer thicknesses in (b) were determined from interference effects observed using in situ and post growth reflectivity. Layer thicknesses in (c) were estimated using growth rates and period of superlattice 3.6~nm known from X-ray diffraction.}
\label{fig:scheme}
\end{figure}

The structure of the low refractive index superlattice MgSe/ZnTe/MgTe/ZnTe is shown in Fig.~\ref{fig:scheme}(c). In the superlattice, layers of a few nanometer thickness with preferential rock salt structure (such as MgSe or MgTe) can be effectively stabilized in zinc blende structure when they are deposited on a zinc blende substrate and embedded between zinc blende layers like ZnTe. It is important to note that both MgSe and MgTe exhibit a refractive index much smaller than ZnTe [both have an energy gap larger than ZnTe, see Fig.~\ref{fig:XRD}(b)].

MgSe has a lattice parameter smaller than ZnTe, MgTe larger than ZnTe [Fig.~\ref{fig:XRD}(b)]. We compensate lattice mismatch by using a defined thickness ration of MgSe and MgTe layers [Fig.~\ref{fig:scheme}(c)]. This is quite practical for epitaxy, because controlling a layer thickness is much easier than balancing strain by controlling the content of a ternary or quaternary compound. MgSe and MgTe layers are separated by very thin ZnTe layers, which are neutral with regard to strain and lattice parameter. Since a ZnTe interlayer undesirably increases the refractive index of such a superlattice structure, it should be kept as thin as possible just to be able to support growth of superlattice in zinc blende structure.


\begin{figure}[bt]
\includegraphics[width=\linewidth]{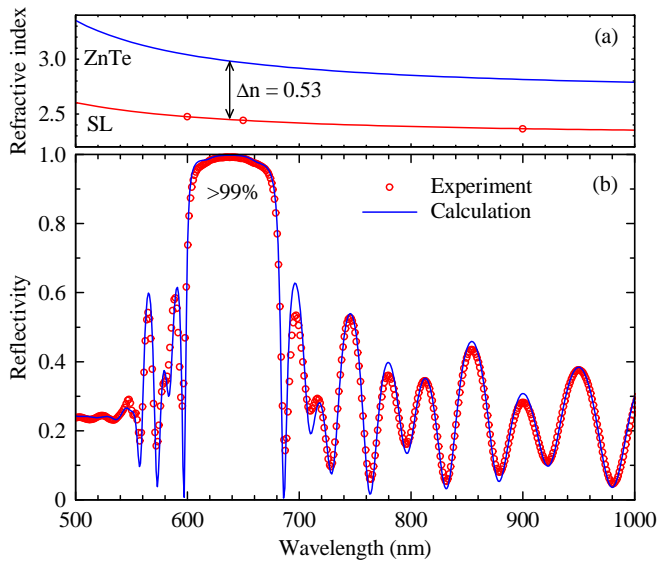}
\centering \caption[]{(color online). (a) Refractive index of ZnTe and triple superlattice (SL) MgSe/ZnTe/MgTe/ZnTe. Points corresponds to the refractive index of SL determined in this work. Curves are calculated using Eq.~\ref{eq:n2} with parameters of Ref.~\onlinecite{Marple64} for ZnTe and with fitting parameters determined in this work for the SL. Note a significant difference between refractive indices of ZnTe and SL, about 0.5 for range of interest marked by arrow. (b) Measured (points) and calculated (solid line) reflectivity spectra of DBR structure lattice matched to ZnTe. There are 15 DBR pairs in the structure. At the center of the stop-band ($\lambda_0=638$~nm) reflectivity coefficient exceeds 99\%.}
\label{fig:reflectivity}
\end{figure}


Our DBR structure was grown using molecular beam epitaxy (MBE). We controlled layer thickness by in-situ reflectivity,\cite{Kruse02} which allowed us to grow quarter-wave thick layer with precision of 1~nm. In this work we used a GaAs substrate covered by a thin ZnSe layer, followed by a thick ZnTe buffer. X-ray diffraction reveals that the top of the ZnTe buffer is completely relaxed, so it has lattice parameter of bulk ZnTe. Next, there are 15 pairs of low and high refractive index layers. They are well resolved in Fig.~\ref{fig:scheme}(a), which is a cross-section image obtained with scanning electron microscopy (SEM). The internal structure of the superlattice (low index layer) cannot by resolved using SEM, but it can be studied using high resolution X-ray diffraction (HRXRD).

Fig.~\ref{fig:XRD}(a) shows a mapping in a range of the reciprocal space close to (335) X-ray reflex. There are intensity maxima related to unstrained GaAs and ZnTe. The intensity maximum related to ZnSe shows that ZnSe is pseudomorphically grown on GaAs (the same Qx position as GaAs). Most interesting are the fringes related to the short period superlattice. They are vertically aligned to the spot of the relaxed ZnTe buffer layer, i.e. the strain in the superlattice is sufficiently balanced to avoid relaxation and to keep pseudomorphic growth of the whole DBR. The spacing between fringes corresponds to a superlattice period length of $3.6\pm0.1$~nm.

The reflectivity spectrum of the DBR is shown in Fig.~\ref{fig:reflectivity}(b). Points denote experimental data and the solid line represents a simulation. With only one fitting parameter, i.e. the center of the stop-band, we observe an almost perfect agreement of experiment and simulation. The highest reflectivity, over 99\%, is observed for $\lambda_0=638$~nm. The observation of such a high value of reflectivity coefficient is an evidence of the good crystalline quality of the structure. The width of the stop-band, over 60 nm, results from a rather high refractive index step $\Delta n$ of about  $0.5$.


The reflectivity spectrum represented by the solid line in Fig.~\ref{fig:reflectivity}(b) was calculated as follows. We use the refractive index of ZnTe described by Marple:\cite{Marple64}

\begin{equation}
n^2 = A + B\lambda^2/(\lambda^2 - c^2),
\label{eq:n2}
\end{equation}

where $n$ is refractive index, $\lambda$ is wavelength in micrometers; $A$, $B$,  and $c^2$ are empirical parameters characteristic for each semiconductor. Values of these parameters for ZnTe are given in Ref.~\onlinecite{Marple64}: $A_{ZnTe} = 4.27$, $B_{ZnTe} = 3.01$,  and $c^2_{ZnTe} = 0.142$. The corresponding dispersion is shown by the upper curve in Fig.~\ref{fig:reflectivity}(a).

The short period triple superlattice MgSe/ZnTe/MgTe/ZnTe is treated as a homogeneous layer with a single dispersion. In order to determine the refractive index $n_{SL}$ of the superlattice, a layer with thickness $d=\lambda/n_{SL}$ was grown using in situ time-dependent reflectivity measurement and $d$ was obtained using the superlattice period length measured by HRXRD. We determined experimental values of $n_{SL}$ for three selected wavelengths, as it is shown by points in Fig.~\ref{fig:reflectivity}(a). Using the same model (Eq. \ref{eq:n2}) and the best fit to the experimental data, we obtained the lower curve of Fig.~\ref{fig:reflectivity}(a) and following values of parameters: $A_{SL} = 3.45$, $B_{SL} = 1.85$,  and $c^2_{SL} = 0.111$. Note that the refractive index of the SL is smaller than that of ZnTe. Moreover, the spectral dependence of refractive index is weaker for the SL compared to ZnTe. Both properties indicate that the SL acts like a semiconductor with an energy gap significantly larger than the energy gap of ZnTe.

Additionally, we included in the model the effect of absorption for photon energies above the energy gap of ZnTe, i.e. for wavelength close to 550 nm and shorter. We used the imaginary part of refractive index of ZnTe given in Ref.~\onlinecite{Sato92}, refractive index of ZnSe given in Ref.~\onlinecite{Marple64} and refractive index of GaAs given in Ref.~\onlinecite{zollner01}. We assumed an  infinite thickness of the GaAs substrate and optimal thickness of each layer in the structure: $d=\lambda_0/(4n_{ZnSe})$ for ZnSe layer, $d=19\lambda_0/(4n_{ZnTe})$ for ZnTe buffer, $d=\lambda_0/(4n_{SL})$ for layers with low refractive index, and $d=\lambda_0/(4n_{ZnTe})$ for layers with high refractive index. These layer thicknesses result in constructive interference observed in the reflected light with the wavelength $\lambda_0$. The parameter $\lambda_0=638$~nm denotes the center wavelength of the stopband. Using these parameters, the reflectivity spectrum was calculated by the transfer matrix method and a good agreement between the experiment and the calculation was found. This indicates that the layer thicknesses are close to the assumed optimal case. The calculated thickness of each layer is shown in Fig.~\ref{fig:scheme}(b).


In conclusion, a high-reflectivity and broad-stopband distributed Bragg reflector lattice matched to ZnTe has been realized. The structure is based on an original approach for the low refractive index material, namely a triple MgSe/ZnTe/MgTe/ZnTe superlattice containing only binary compounds. The reflectivity coefficient of our DBR exceeds 99\%. The reflectivity spectrum is well reproduced by a simple model based on the transfer matrix method and the experimentally determined dispersion of the superlattice. This DBR will be a key component for the fabrication of microcavities with a high quality factor in the future, paving the way for efficient emitters in the green to orange spectral range and applications of spin-related quantum information science.


This work was supported by Marie Curie Actions (contract number	MTKD-CT-2005-029671), Foundation for Polish Science, Deutscher Akademischer Austausch Dienst, and Alexander von Humboldt Foundation. The authors gratefully acknowledge skillful assistance by Kai Otte and Torben Rohbeck.


\end{document}